\documentclass[aps,prd,preprintnumbers,superscriptaddress,nofootinbib,twocolumn]{revtex4-1}
\usepackage[pdftex]{graphicx}
\usepackage{bm,latexsym,amsmath,amssymb,amsfonts,mathrsfs}
\allowdisplaybreaks[1]
\newcommand*{\D}{{\rm d}}

\begin{document}
\title{On the screening mechanism in DHOST theories evading gravitational wave constraints}
\author{Shin'ichi~Hirano}
\email[Email: ]{s.hirano"at"rikkyo.ac.jp}
\affiliation{Department of Physics, Rikkyo University, Toshima, Tokyo 171-8501, Japan
}

\author{Tsutomu~Kobayashi}
\email[Email: ]{tsutomu"at"rikkyo.ac.jp}
\affiliation{Department of Physics, Rikkyo University, Toshima, Tokyo 171-8501, Japan
}

\author{Daisuke~Yamauchi}
\email[Email: ]{yamauchi"at"jindai.jp}
\affiliation{Faculty of Engineering, Kanagawa University, Kanagawa, 221-8686, Japan}

\begin{abstract}
We consider a subclass of degenerate higher-order scalar-tensor (DHOST) theories
in which gravitational waves propagate at the speed of light
and do not decay into scalar fluctuations.
The screening mechanism in DHOST theories evading these two gravitational wave constraints
operates very differently from that in generic DHOST theories.
We derive
a spherically symmetric solution in the presence of
nonrelativistic matter. General relativity is
recovered in the vacuum exterior region
provided that functions in the Lagrangian satisfy a certain condition,
implying that
fine-tuning is required.
Gravity in the matter interior exhibits novel features:
although the gravitational potentials still obey the standard inverse power law,
the effective gravitational constant is different from its exterior value,
and the two metric potentials
do not coincide.
We discuss possible observational constraints on this subclass of DHOST theories,
and argue that the tightest bound comes from the Hulse-Taylor pulsar.
\end{abstract}
\pacs{%
04.50.Kd  
}
\preprint{RUP-19-8}
\maketitle
\section{Introduction}

Measuring the speed of gravitational waves serves as a
powerful test for modified theories of
gravity~\cite{Nishizawa:2014zna,Lombriser:2015sxa,Lombriser:2016yzn,Bettoni:2016mij}.
Based on this idea,
the nearly simultaneous detection of the gravitational wave event
GW170817 and its electromagnetic counterpart
GRB 170817A~\cite{TheLIGOScientific:2017qsa,Monitor:2017mdv,GBM:2017lvd}
was used to put a tight limit on scalar-tensor
theories as alternatives to dark
energy~\cite{Creminelli:2017sry,Sakstein:2017xjx,Ezquiaga:2017ekz,Baker:2017hug,%
Bartolo:2017ibw,Kase:2018iwp,Ezquiaga:2018btd,Kase:2018aps}.\footnote{The constraints
have been imposed assuming that modified gravity as an alternative to dark energy
is valid on much higher energy scales where
LIGO observations are made, though this assumption may be subtle~\cite{deRham:2018red}.}
Within the Horndeski class of scalar-tensor
theories~\cite{Horndeski:1974wa,Deffayet:2011gz,Kobayashi:2011nu},
derivative couplings of the scalar degree of freedom $\phi$
to the curvature have thus been ruled out.
One can extend the Horndeski theory
in a healthy manner to
degenerate higher-order scalar-tensor (DHOST) theories,
in which Ostrogradsky instabilities are eliminated
despite the higher-order Euler-Lagrange
equations~\cite{Zumalacarregui:2013pma,Gleyzes:2014dya,Gleyzes:2014qga,%
Langlois:2015cwa,Langlois:2015skt,Crisostomi:2016czh,%
Achour:2016rkg,BenAchour:2016fzp,Langlois:2017mxy}
(see Refs.~\cite{Langlois:2017mdk,Langlois:2018dxi,Kobayashi:2019hrl}
for a review).
Nontrivial derivative couplings to the curvature are
still allowed in the context of DHOST theories.
These theories are phenomenologically very interesting
because while the Vainshtein screening mechanism is
successfully implemented in the vacuum region exterior to
matter distributions, it is partially broken
in the matter interior~\cite{Kobayashi:2014ida,Crisostomi:2017lbg,Langlois:2017dyl,Dima:2017pwp}.
This implies that DHOST theories can only be tested in the interior of
extended objects such as stars, galaxy clusters, and Earth's
atmosphere~\cite{Koyama:2015oma,Saito:2015fza,%
Sakstein:2015zoa,Sakstein:2015aac,Jain:2015edg,%
Sakstein:2016ggl,Sakstein:2016lyj,Salzano:2017qac,Saltas:2018mxc,
Babichev:2016jom,Sakstein:2016oel,Chagoya:2018lmv,Kobayashi:2018xvr,
Babichev:2018rfj}.

Recently, yet another constraint on DHOST theories has been
pointed out:
gravitons must be stable against
decay into dark energy~\cite{Creminelli:2018xsv}.
The Lagrangian for DHOST theories
in which gravitons propagate at the speed of light and do not decay
into dark energy is described by
\begin{align}
{\cal L}&=G_2(\phi,X)-G_3(\phi,X)\Box\phi
\notag \\ & \quad
+f(\phi,X){\cal R}+\frac{3f_X^2}{2f}
\phi^\mu\phi_{\mu\sigma}\phi^{\sigma \nu}\phi_\nu,
\label{eq:dhostlag}
\end{align}
where ${\cal R}$ is the Ricci scalar, $\phi_\mu=\nabla_\mu\phi$,
$\phi_{\mu\nu}=\nabla_\mu\nabla_\nu\phi$, 
$X:=-\phi_\mu\phi^\mu/2$, and $f_X=\partial f/\partial X$.
Cosmology derived from the Lagrangian~\eqref{eq:dhostlag}
is explored in Ref.~\cite{Frusciante:2018tvu}.
It turns out that in this particular subclass of DHOST theories
the screening mechanism operates in a different way
from that in generic DHOST theories, as already inferred in Ref.~\cite{Creminelli:2018xsv}.
The purpose of the present paper is to clarify
how the (breaking of the) Vainshtein screening mechanism occurs
in the above theory.

\section{Screening mechanism in DHOST theories without graviton decay}

A weak gravitational field is described by the line element
\begin{align}
\D s^2=-[1+2\Phi(t,\Vec{x})]\D t^2+[1-2\Psi(t,\Vec{x})]\D \Vec{x}^2,
\end{align}
with the scalar-field configuration
\begin{align}
\phi = \phi_0(t) + \pi (t,\Vec{x}).
\end{align}
Here, $\phi_0(t)$ is a slowly evolving background
determined from the cosmological boundary condition
and $\pi(t,\Vec{x})$ is a fluctuation.
Since we are interested in gravity on scales well inside the horizon,
we ignore the cosmic expansion.

Following Refs.~\cite{Koyama:2013paa,Kobayashi:2014ida}, we expand the action
in terms of the metric perturbations and $\pi$,
keeping the higher-derivative terms relevant to
the screening mechanism in the quasi-static regime.
The resultant effective Lagrangian is given by
\begin{align}
{\cal L}_{\rm eff}&=f\biggl[
-2\Psi\partial^2\Psi+4(1-2\beta)\Psi\partial^2\Phi -\frac{\eta}{2f}(\partial\pi)^2
\notag \\ & \quad
+4\beta\left(1-\frac{3\beta}{2}\right)\Phi\partial^2\Phi
+\frac{4\xi}{f^{1/2}}\Psi\partial^2\pi
\notag \\ & \quad
+\frac{2(\alpha-\xi)}{f^{1/2}}
\Phi\partial^2\pi
+\frac{\alpha}{f \Lambda^3}(\partial\pi)^2\partial^2\pi
\notag \\ & \quad
+\frac{2\beta\left(1-3\beta\right)}{f^{1/2}\Lambda^3}(\partial\pi)^2\partial^2\Phi
-\frac{4\beta}{f^{1/2}\Lambda^3}(\partial\pi)^2\partial^2\Psi
\notag \\ &\quad
+\frac{6\beta^2}{f\Lambda^6}\partial_i\pi \partial_j\pi\partial_i\partial_k \pi
\partial_k\partial_j \pi
\notag \\ & \quad
+\frac{6\beta^2}{f^{1/2}\Lambda^3}(\partial\dot\pi)^2
-\frac{4\beta(1-3\beta)\dot\phi_0}{f^{1/2}\Lambda^3}\Phi\partial^2\dot\pi
\notag \\ & \quad
+\frac{8\beta\dot\phi_0}{f^{1/2}\Lambda^3}\Psi\partial^2\dot\pi
+\frac{6\beta^2\dot\phi_0}{f\Lambda^6}(\partial\pi)^2\partial^2\dot\pi
\biggr]
 -\Phi \rho,\label{effLag}
\end{align}
where we introduced dimensionless quantities
\begin{align}
\alpha:= \frac{\dot\phi_0^2G_{3X}}{2f^{1/2}},\quad
\beta:=\frac{\dot\phi_0^2f_X}{2f}, \quad \xi:=\frac{f_\phi}{f^{1/2}},
\end{align}
and defined
an energy scale $\Lambda:=(\dot\phi_0^2/f^{1/2})^{1/3}$.
The dot denotes differentiation with respect to $t$.
The explicit expression for the coefficient $\eta$ is not important here.
In deriving the Lagrangian~\eqref{effLag} we
ignored $\ddot\phi_0$ since $\phi_0$ is a slowly varying field.
We assume that matter is minimally coupled to gravity,
so that
we add the term $-\Phi \rho$
where $\rho=\rho(t,\Vec{x})$
is the density of a nonrelativistic matter source.
The Lagrangian~\eqref{effLag} is a particular case
of the general effective Lagrangian for the Vainshtein mechanism
in DHOST theories~\cite{Crisostomi:2017lbg,Langlois:2017dyl,Dima:2017pwp}.
However, the screening mechanism
in this particular subclass
operates in a very different way
than in generic cases, as we will see below.

Let us consider a spherically symmetric matter distribution,
$\rho=\rho(t,r)$,
where $r$ is the radial coordinate. Varying the action with respect to
$\Psi$, $\Phi$, and $\pi$, we obtain the following equations:
\begin{align}
&(1-\beta)\xi x+(1-2\beta)y-z-2\beta x(rx)'
+\frac{2\dot\phi_0}{\Lambda^3}\beta\dot x=0,\label{psieq}
\\
&[\alpha-\xi+(1-3\beta)\beta \xi]x
+2\beta(2-3\beta)y+2(1-2\beta)z
\notag \\ &
+2\beta(1-3\beta)x(rx)'
-\frac{2\dot\phi_0}{\Lambda^3}\beta(1-3\beta)\dot x
=A,\label{phieq}
\end{align}
and
\begin{align}
{\cal F}(x,\dot x,x',\ddot x,\dot x',x'',y,\dot y, y', z, \dot z, z')=0,
\label{pieq}
\end{align}
where
the prime denotes differentiation with respect to $r$ and
we defined the dimensionless variables as
\begin{align}
&x:=\frac{\pi'}{\Lambda^3r},\quad
y:=\frac{f^{1/2}\Phi'}{\Lambda^3r},\quad z:=\frac{f^{1/2}\Psi'}{\Lambda^3r},
\\ &
A:=
\frac{1}{8\pi \dot\phi_0^2}\frac{M(t,r)}{r^3}=
\frac{1}{8\pi f^{1/2}\Lambda^3}\frac{M(t,r)}{r^3},
\end{align}
with
\begin{align}
M(t,r):=4\pi \int^r_0\rho(t,\bar r)\bar r^2\D \bar r
\end{align}
being the mass contained within $r$.
In deriving Eqs.~\eqref{psieq}--\eqref{pieq} we integrated
the field equations once and fixed the integration constants
so that $x$, $y$, and $z$ are regular at $r=0$.
The explicit form of ${\cal F}$ is complicated.

From Eqs.~\eqref{psieq} and~\eqref{phieq} we have
\begin{align}
y&=\frac{A+2\beta(1-\beta)x(rx)'}{2(1-\beta)^2}+c_1 x
-\frac{\dot\phi_0}{\Lambda^3}
\frac{\beta}{1-\beta}\dot x,
\label{eq:y=}
\\
z&=\frac{(1-2\beta)A-2\beta(1-\beta)x(rx)'}{2(1-\beta)^2}
+c_2x +\frac{\dot\phi_0}{\Lambda^3}
\frac{\beta}{1-\beta}\dot x,\label{eq:z=}
\end{align}
where $c_1$ and $c_2$ are written in terms of $\alpha$, $\beta$, and $\xi$.
Then, substituting Eqs.~\eqref{eq:y=} and~\eqref{eq:z=} to
Eq.~\eqref{pieq}, we obtain
\begin{align}
&4(\alpha-3\beta\xi)(1-\beta)x^2+\left[c_3
-2\beta(1-\beta)\frac{(r^3A)'}{r^2}
\right]x
\notag \\ &
=[\alpha+(1-2\beta)\xi-2\zeta]A
-\frac{2\dot\phi_0}{\Lambda^3}(1-\beta)\beta \dot A,\label{eq:x2}
\end{align}
where we defined
\begin{align}
\zeta:= \frac{\dot\phi_0^2f_{\phi X}}{2f^{1/2}},
\end{align}
and the explicit expression for $c_3$
(which is written in terms of $\alpha$, $\beta$, etc. and their time derivatives)
is not important.
As expected from the degeneracy of the theory,
the final result~\eqref{eq:x2} is just an algebraic equation for $x$,
with no derivatives acting on $x$.
In generic quadratic DHOST theories, however,
one would obtain at this final stage a cubic equation for $x$.
The present theory is special in the sense that
the coefficient of the cubic term vanishes identically.

From now on, let us consider the case where the source is static,
$\rho=\rho(r)$. Then, since we are assuming that $\dot\phi_0$
is approximately constant, $A$ is also independent of time.
Thus, $\dot A$ in Eq.~\eqref{eq:x2} can be neglected.

One may define the typical radius $r_V$
below which nonlinearities are large
by $A(r_V)=1$.
We are mainly interested in the solutions to Eq.~\eqref{eq:x2}
for $A\gg 1$ both inside and outside the matter source.
Outside the matter distribution we have $A\propto r^{-3}$,
whereas we have $(r^3A)'\neq 0$ inside.

Let us first consider the exterior region.
For $A\gg 1$ we have
\begin{align}
x\simeq  \pm\frac{1}{2}
\left[\frac{\alpha+(1-2\beta)\xi-2\zeta}{(\alpha-3\beta\xi)(1-\beta)}A \right]^{1/2}.
\label{gg1xsol}
\end{align}
From this it can be seen
that the terms linear in $x$ in Eqs.~\eqref{eq:y=} and~\eqref{eq:z=}
are suppressed relative to the other terms.
We thus find,
irrespective of the sign of Eq.~\eqref{gg1xsol}, that
\begin{align}
y&\simeq
\frac{\alpha(4-\beta)-\beta(13-2\beta)\xi+2\beta\zeta}{8(\alpha-3\beta\xi)(1-\beta)^2}A,
\\
z&\simeq
\frac{\alpha(4-7\beta)-11\beta(1-2\beta)\xi-2\beta\zeta}{8(\alpha-3\beta\xi)(1-\beta)^2}A,
\end{align}
This shows that $\Phi\neq \Psi$ in general,
implying that the present subclass of DHOST theories
does not evade the solar-system constraints.
However, if the parameters satisfy\footnote{More precisely,
the condition for successful screening is
$\beta[3\alpha-\xi(1+10\beta)+2\zeta]=0$. Clearly, the case with $\beta=0$
corresponds to the subclass of the Horndeski theory. This is the trivial
case exhibiting the Vainshtein
mechanism~\cite{Kimura:2011dc,Narikawa:2013pjr,Koyama:2013paa}.}
\begin{align}
3\alpha-\xi(1+10\beta)+2\zeta=0,\label{out_relation}
\end{align}
general relativity is recovered, yielding
\begin{align}
&y=z=\frac{A}{2(1-\beta)},\notag\\
\quad\Leftrightarrow\quad &\Phi' = \Psi' = \frac{1}{16\pi f(1-\beta)}\frac{M}{r^2}.
\end{align}
The effective gravitational constant is given by
\begin{align}
G_{N, {\rm out}}=\frac{1}{16\pi f(1-\beta)}.
\end{align}
Thus, 
fine-tuning is needed in order for the
screening mechanism to work successfully in the vicinity of a source.
This is in contrast to generic
DHOST theories~\cite{Kobayashi:2014ida,Crisostomi:2017lbg,Langlois:2017dyl,Dima:2017pwp}.

Next, let us look at the interior region.
We have two branches, one of which is given by
\begin{align}
{\rm (I)}:\quad
x\simeq \frac{\beta}{2(\alpha-3\beta\xi)}
\frac{(r^3A)'}{r^2}\gg 1,
\end{align}
and the other by
\begin{align}
  {\rm (II)}:\quad
x\simeq -\frac{\alpha+(1-2\beta)\xi-2\zeta}{2\beta(1-\beta)}
\frac{r^2A}{(r^3A)'}={\cal O}(1).
\end{align}

In Branch I,
the behavior of gravity
is far away from the
normal one:
\begin{align}
y&=  \frac{9\beta^3}{(1-\beta)^3\xi^2}
\frac{(r^3A)'}{r^3}\left[
(r^3A)''-\frac{(r^3A)'}{r}
\right]+{\cal O}(A),\label{ybranch1}
\\
z &= -\frac{9\beta^3}{(1-\beta)^3\xi^2}
\frac{(r^3A)'}{r^3}\left[
(r^3A)''-\frac{(r^3A)'}{r}
\right]+{\cal O}(A),\label{zbranch1}
\end{align}
where Eq.~\eqref{out_relation} was assumed.
It then follows that
\begin{align}
\Phi'\simeq -\Psi' \propto \frac{M'M''}{r^2}-\frac{(M')^2}{r^3}.
\label{branch_1}
\end{align}
We therefore conclude that
this branch 
would not describe the stellar structure appropriately,
and hence must be excluded.

Branch II is phenomenologically more interesting.
In this branch, all $x$'s in Eqs.~\eqref{eq:y=} and~\eqref{eq:z=}
 can be neglected, leading to
\begin{align}
&y= \frac{A}{2(1-\beta)^2},
\quad
z= \frac{(1-2\beta )A}{2(1-\beta)^2}.
\notag\\
\quad\Leftrightarrow\quad
&\Phi' = \frac{1}{16\pi f(1-\beta)^2}\frac{M}{r^2},
\quad
\Psi = (1-2\beta)\Phi.\label{intresult}
\end{align}
From this we see that the effective gravitational constant
inside the matter distribution is different from the exterior value
by a factor of $(1-\beta)^{-1}$:
\begin{align}
G_{N,{\rm in}}=\frac{G_{N,{\rm out}}}{1-\beta}.
\end{align}
This must be contrasted with the way of
breaking the screening mechanism in generic DHOST theories,
where $M'$ and $M''$ appear in $\Phi'$ and $\Psi'$
as corrections to the standard gravitational law with the same
gravitational constant as the exterior
one~\cite{Kobayashi:2014ida,Crisostomi:2017lbg,Langlois:2017dyl,Dima:2017pwp}.
We also see that $\Phi$ and $\Psi$ do not coincide
in the matter interior.
One should note that Eq.~\eqref{out_relation}
is not used when deriving Eq.~\eqref{intresult}.

Let us finally comment on the solution for $A\ll 1$.
We have two branches, namely, $x\sim y\sim z \sim A$ and $x\sim y\sim z\sim 1$.
By inspecting the explicit solutions to Eq.~\eqref{eq:x2},
we find that the former branch, which
is phenomenologically more acceptable, is matched onto
Branch II if
\begin{align}
\beta(1-\beta)c_3<0
\end{align}
is satisfied.

\begin{figure}[tb]
    \includegraphics[keepaspectratio=true,height=50mm]{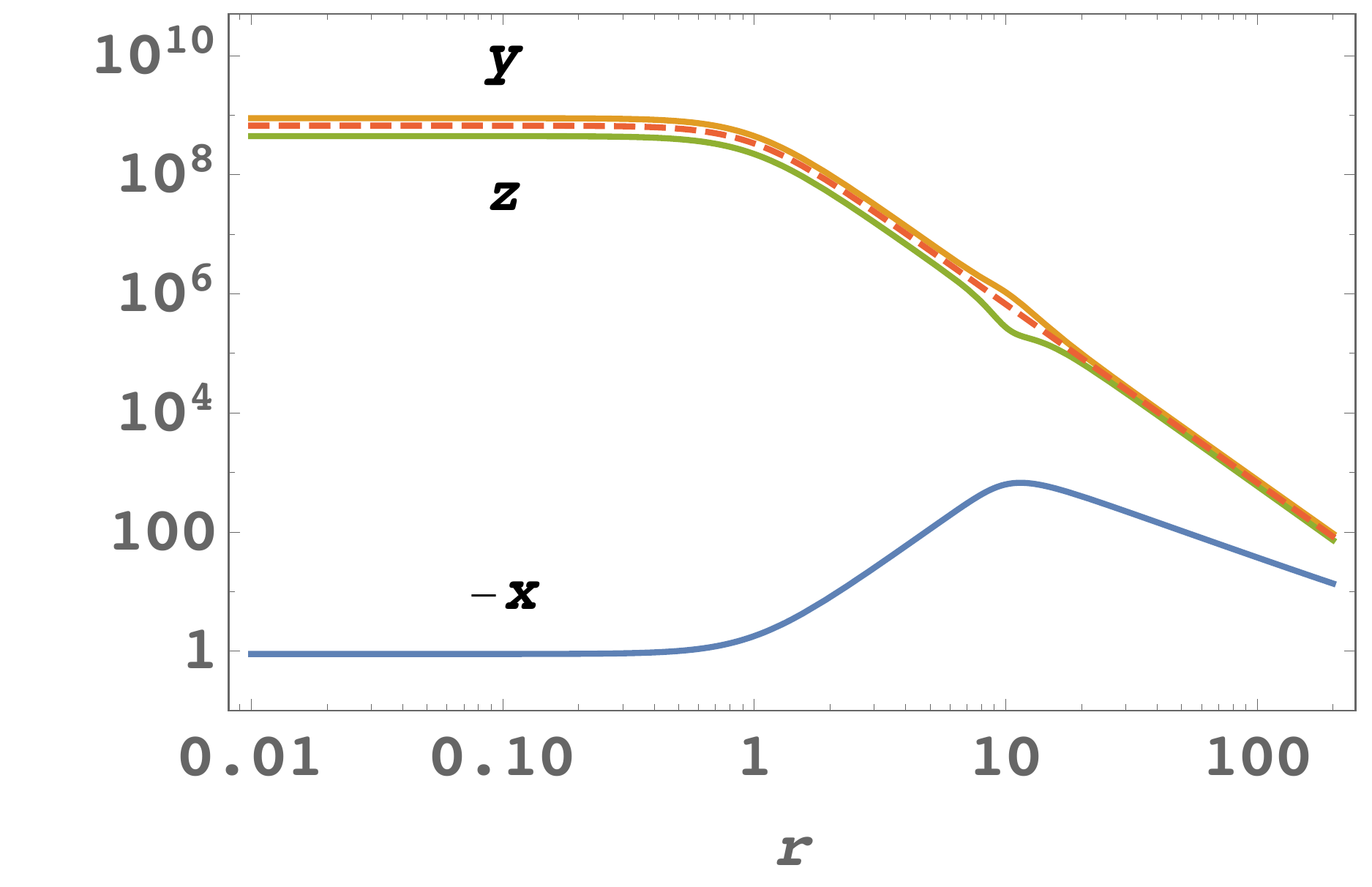}
     \caption{An example of a Branch II solution for $r_V=1000$
     and the stellar radius $\sim 1$. The dashed line
     corresponds to the potentials in GR with the gravitational
      constant $G_{N,{\rm out}}$.
	}
     \label{fig:1}
\end{figure}

\begin{figure}[tb]
    \includegraphics[keepaspectratio=true,height=50mm]{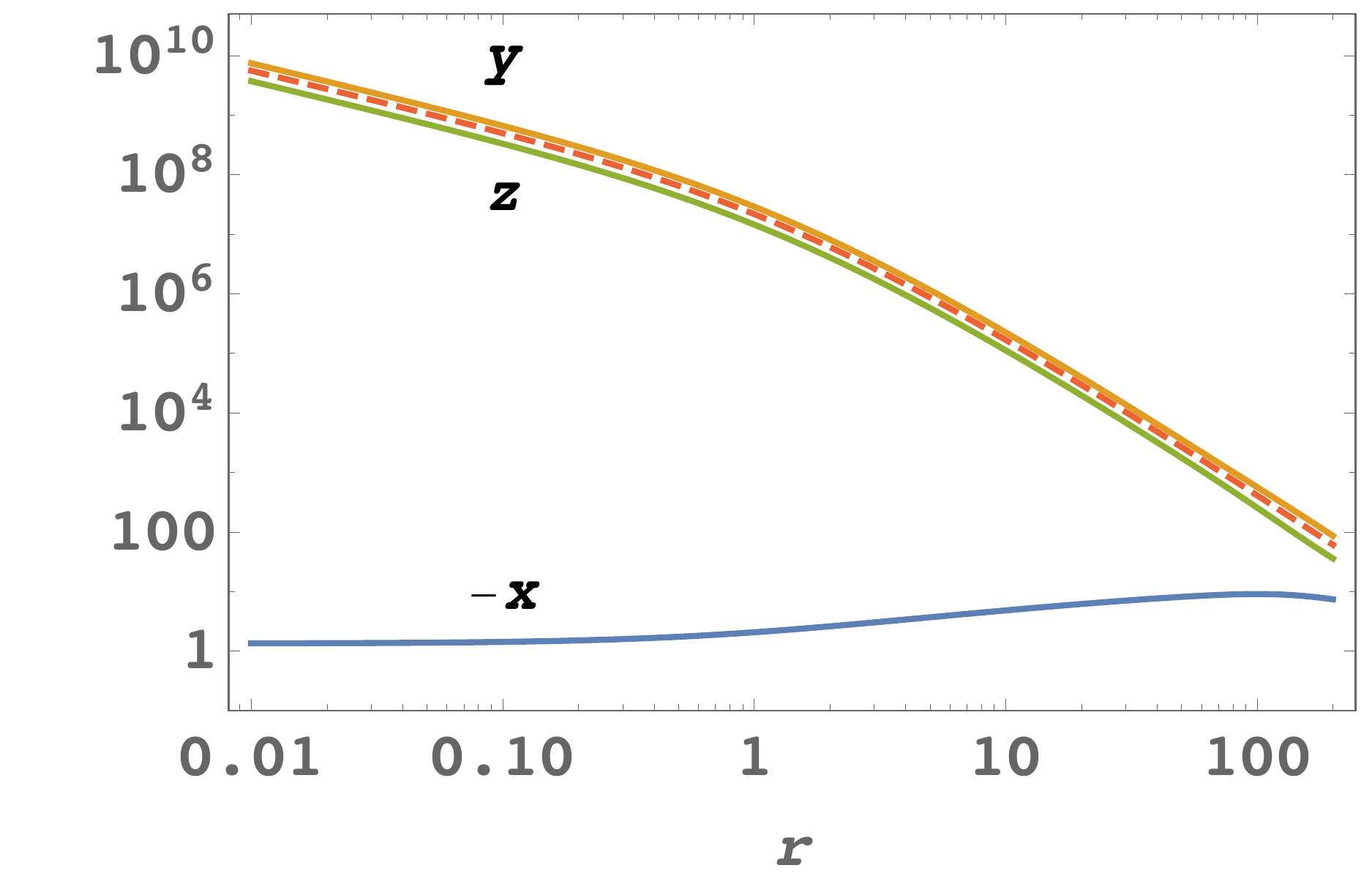}
     \caption{The Branch II solution for
the NFW density profile.
The dashed line
     corresponds to the potentials in GR with the gravitational
      constant $G_{N,{\rm out}}$.
	}
     \label{fig:2}
\end{figure}

As an example, we show in Fig.~\ref{fig:1}
the Branch II profiles of $x$, $y$, and $z$ for
$A(r)=B(r)/B(1000)$ (namely, $r_V=1000$) with $B(r)=(r^3+1)^{-1}$.
The density profile mimics a star
with the radius $r\sim 1$. The parameters are given by
$\xi=\alpha=1$, $\beta=\zeta=1/4$, and $c_3=1$.
(For $x$ we plot an exact solution to Eq.~\eqref{eq:x2}, but
for $y$ and $z$ the terms linear in $x$ are ignored
because they are subdominant for $r\ll r_V$.)

We also present in Fig.~\ref{fig:2}
the Branch II solution for the NFW density profile,
$\rho(r)=\rho_0/[(r/r_s)(1+r/r_s)^2]$
with $r_s=1$ and $\rho_0$ chosen so that $r_V=1000$.
The parameters are again given by
$\xi=\alpha=1$, $\beta=\zeta=1/4$, and $c_3=1$.
Since there is no definite surface in this case,
we see deviations from general relativity everywhere.

\section{Observational constraints}

We have seen that though the particular subclass of DHOST theories~(\ref{eq:dhostlag})
could evade solar-system tests by requiring
the fine-tuned relation~\eqref{out_relation},
(i)
$\Phi$ and $\Psi$ do not coincide inside the matter distribution, and
(ii)
the gravitational constant in the matter interior is
different from its exterior value.
Let us discuss briefly possible observational constraints
on such modifications of gravity.

The difference between the two potentials in the nonvacuum region,
$\Psi/\Phi-1=-2\beta$,
can be measured
by comparing the X-ray and lensing profiles of galaxy clusters,
as has been investigated for different types of modifications
in Refs.~\cite{Terukina:2013eqa,Wilcox:2015kna,Sakstein:2016ggl}.
In particular,
the constraints obtained for beyond Horndeski theories
in Ref.~\cite{Sakstein:2016ggl}
read
$|\Phi/\Phi_{\rm GR}-1|< {\cal O}(10^{-1})$
and
$|\Psi/\Psi_{\rm GR}-1|< {\cal O}(10^{-1})$.
Thus, we would expect
constraints of the same order of magnitude,
$|\beta|<{\cal O}(10^{-1})$, from galaxy clusters.

A different value of the gravitational constant inside the Sun
would lead to changes in the solar structure, and thereby
modify the sound speed and solar neutrino fluxes.
Based on the solar standard model, it has been argued that
a relative difference of ${\cal O}(10^{-2})$
is still allowed by observations~\cite{Lopes:2003aa}.
Thus, the Sun could potentially be used to test
a different value of the gravitational constant
inside extended objects.

Note, however, that currently
the most stringent bound comes from
the difference between the measured value of the
gravitational constant,
$G_N$($=G_{N,{\rm out}}$ or $G_{N,{\rm in}}$),
and the gravitational coupling for gravitational waves, $G_{\rm GW}$,
which is constrained from
the orbital decay of the Hulse-Taylor pulsar:
$-7.5\times 10^{-3} <%
G_{\rm GW}/G_N-1<2.5\times 10^{-3}$~\cite{Jimenez:2015bwa,Dima:2017pwp}.
In the present case, we have
$G_{\rm GW}=(16\pi f)^{-1}$~\cite{deRham:2016wji,Langlois:2017mxy},
so the constraint is given by
\begin{align}
|\beta|<{\cal O}(10^{-3}),
\end{align}
which is orders of magnitude
tighter than the possible constraint from galaxy clusters.


\section{Conclusions}

In this paper, we have studied the screening mechanism
in a particular subclass of degenerate higher-order scalar-tensor (DHOST) theories
in which the speed of gravitational waves is equal to
the speed of light and gravitons do not decay into scalar fluctuations.
By inspecting a spherically symmetric gravitational field,
we have found that the screening mechanism operates in a
very different way from that in generic DHOST
theories~\cite{Kobayashi:2014ida,Crisostomi:2017lbg,Langlois:2017dyl,Dima:2017pwp}.
First,
the fine-tuning is required so that
solar-system tests are evaded in the vacuum exterior region.
This is in contrast to generic DHOST theories,
in which the implementation of the Vainshtein screening mechanism
outside the matter distribution is rather automatic.
Second, the way of the Vainshtein breaking inside
extended objects is also different from that in generic DHOST theories.
We have shown that in the interior region
the metric potentials
obey the
standard inverse power law, but the two do not coincide.
Moreover, the effective gravitational constant differs from its
exterior value.
However, the current most stringent bound comes from
the fact that the effective gravitational coupling for gravitational waves
is different from the Newtonian constant~\cite{Jimenez:2015bwa,Dima:2017pwp},
rather than from the above interesting phenomenology.
The obtained constraint is as tight as
\begin{align}
\left|\frac{Xf_X}{f}\right| <{\cal O}(10^{-3}).
\end{align}
Thus, we conclude that
the allowed parameter space is small for
DHOST theories as alternatives to dark energy
evading gravitational wave constraints.

\acknowledgments
The work of SH was supported by the JSPS Research Fellowships for Young Scientists
No.~17J04865.
The work of TK was supported by
MEXT KAKENHI Grant Nos.~JP15H05888, JP17H06359, JP16K17707, JP18H04355,
and MEXT-Supported Program for the Strategic Research Foundation at Private Universities,
2014-2018 (S1411024).
The work of DY was supported by MEXT KAKENHI Grant No.~JP17K14304.


\bibliography{A3}

\providecommand{\href}[2]{#2}\begingroup\raggedright\begin{thebibliography}{10}

\bibitem{Nishizawa:2014zna}
A.~Nishizawa and T.~Nakamura, \emph{{Measuring Speed of Gravitational Waves by
  Observations of Photons and Neutrinos from Compact Binary Mergers and
  Supernovae}}, \href{https://doi.org/10.1103/PhysRevD.90.044048}{\emph{Phys.
  Rev.} {\bfseries D90} (2014) 044048}
  [\href{https://arxiv.org/abs/1406.5544}{{\ttfamily 1406.5544}}].

\bibitem{Lombriser:2015sxa}
L.~Lombriser and A.~Taylor, \emph{{Breaking a Dark Degeneracy with
  Gravitational Waves}},
  \href{https://doi.org/10.1088/1475-7516/2016/03/031}{\emph{JCAP} {\bfseries
  1603} (2016) 031} [\href{https://arxiv.org/abs/1509.08458}{{\ttfamily
  1509.08458}}].

\bibitem{Lombriser:2016yzn}
L.~Lombriser and N.~A. Lima, \emph{{Challenges to Self-Acceleration in Modified
  Gravity from Gravitational Waves and Large-Scale Structure}},
  \href{https://doi.org/10.1016/j.physletb.2016.12.048}{\emph{Phys. Lett.}
  {\bfseries B765} (2017) 382}
  [\href{https://arxiv.org/abs/1602.07670}{{\ttfamily 1602.07670}}].

\bibitem{Bettoni:2016mij}
D.~Bettoni, J.~M. Ezquiaga, K.~Hinterbichler and M.~Zumalac\'{a}rregui,
  \emph{{Speed of Gravitational Waves and the Fate of Scalar-Tensor Gravity}},
  \href{https://doi.org/10.1103/PhysRevD.95.084029}{\emph{Phys. Rev.}
  {\bfseries D95} (2017) 084029}
  [\href{https://arxiv.org/abs/1608.01982}{{\ttfamily 1608.01982}}].

\bibitem{TheLIGOScientific:2017qsa}
{\scshape LIGO Scientific, Virgo} collaboration, B.~Abbott et~al.,
  \emph{{GW170817: Observation of Gravitational Waves from a Binary Neutron
  Star Inspiral}},
  \href{https://doi.org/10.1103/PhysRevLett.119.161101}{\emph{Phys. Rev. Lett.}
  {\bfseries 119} (2017) 161101}
  [\href{https://arxiv.org/abs/1710.05832}{{\ttfamily 1710.05832}}].

\bibitem{Monitor:2017mdv}
{\scshape LIGO Scientific, Virgo, Fermi-GBM, INTEGRAL} collaboration, B.~P.
  Abbott et~al., \emph{{Gravitational Waves and Gamma-rays from a Binary
  Neutron Star Merger: GW170817 and GRB 170817A}},
  \href{https://doi.org/10.3847/2041-8213/aa920c}{\emph{Astrophys. J.}
  {\bfseries 848} (2017) L13}
  [\href{https://arxiv.org/abs/1710.05834}{{\ttfamily 1710.05834}}].

\bibitem{GBM:2017lvd}
{\scshape LIGO Scientific, Virgo, Fermi GBM, INTEGRAL, IceCube, AstroSat
  Cadmium Zinc Telluride Imager Team, IPN, Insight-Hxmt, ANTARES, Swift, AGILE
  Team, 1M2H Team, Dark Energy Camera GW-EM, DES, DLT40, GRAWITA, Fermi-LAT,
  ATCA, ASKAP, Las Cumbres Observatory Group, OzGrav, DWF (Deeper Wider Faster
  Program), AST3, CAASTRO, VINROUGE, MASTER, J-GEM, GROWTH, JAGWAR,
  CaltechNRAO, TTU-NRAO, NuSTAR, Pan-STARRS, MAXI Team, TZAC Consortium, KU,
  Nordic Optical Telescope, ePESSTO, GROND, Texas Tech University, SALT Group,
  TOROS, BOOTES, MWA, CALET, IKI-GW Follow-up, H.E.S.S., LOFAR, LWA, HAWC,
  Pierre Auger, ALMA, Euro VLBI Team, Pi of Sky, Chandra Team at McGill
  University, DFN, ATLAS Telescopes, High Time Resolution Universe Survey,
  RIMAS, RATIR, SKA South Africa/MeerKAT} collaboration, B.~P. Abbott et~al.,
  \emph{{Multi-messenger Observations of a Binary Neutron Star Merger}},
  \href{https://doi.org/10.3847/2041-8213/aa91c9}{\emph{Astrophys. J.}
  {\bfseries 848} (2017) L12}
  [\href{https://arxiv.org/abs/1710.05833}{{\ttfamily 1710.05833}}].

\bibitem{Creminelli:2017sry}
P.~Creminelli and F.~Vernizzi, \emph{{Dark Energy after GW170817 and
  GRB170817A}},
  \href{https://doi.org/10.1103/PhysRevLett.119.251302}{\emph{Phys. Rev. Lett.}
  {\bfseries 119} (2017) 251302}
  [\href{https://arxiv.org/abs/1710.05877}{{\ttfamily 1710.05877}}].

\bibitem{Sakstein:2017xjx}
J.~Sakstein and B.~Jain, \emph{{Implications of the Neutron Star Merger
  GW170817 for Cosmological Scalar-Tensor Theories}},
  \href{https://doi.org/10.1103/PhysRevLett.119.251303}{\emph{Phys. Rev. Lett.}
  {\bfseries 119} (2017) 251303}
  [\href{https://arxiv.org/abs/1710.05893}{{\ttfamily 1710.05893}}].

\bibitem{Ezquiaga:2017ekz}
J.~M. Ezquiaga and M.~Zumalac\'{a}rregui, \emph{{Dark Energy After GW170817:
  Dead Ends and the Road Ahead}},
  \href{https://doi.org/10.1103/PhysRevLett.119.251304}{\emph{Phys. Rev. Lett.}
  {\bfseries 119} (2017) 251304}
  [\href{https://arxiv.org/abs/1710.05901}{{\ttfamily 1710.05901}}].

\bibitem{Baker:2017hug}
T.~Baker, E.~Bellini, P.~G. Ferreira, M.~Lagos, J.~Noller and I.~Sawicki,
  \emph{{Strong constraints on cosmological gravity from GW170817 and GRB
  170817A}}, \href{https://doi.org/10.1103/PhysRevLett.119.251301}{\emph{Phys.
  Rev. Lett.} {\bfseries 119} (2017) 251301}
  [\href{https://arxiv.org/abs/1710.06394}{{\ttfamily 1710.06394}}].

\bibitem{Bartolo:2017ibw}
N.~Bartolo, P.~Karmakar, S.~Matarrese and M.~Scomparin, \emph{{Cosmic
  structures and gravitational waves in ghost-free scalar-tensor theories of
  gravity}}, \href{https://doi.org/10.1088/1475-7516/2018/05/048}{\emph{JCAP}
  {\bfseries 1805} (2018) 048}
  [\href{https://arxiv.org/abs/1712.04002}{{\ttfamily 1712.04002}}].

\bibitem{Kase:2018iwp}
R.~Kase and S.~Tsujikawa, \emph{{Dark energy scenario consistent with GW170817
  in theories beyond Horndeski gravity}},
  \href{https://doi.org/10.1103/PhysRevD.97.103501}{\emph{Phys. Rev.}
  {\bfseries D97} (2018) 103501}
  [\href{https://arxiv.org/abs/1802.02728}{{\ttfamily 1802.02728}}].

\bibitem{Ezquiaga:2018btd}
J.~M. Ezquiaga and M.~Zumalac\'{a}rregui, \emph{{Dark Energy in light of
  Multi-Messenger Gravitational-Wave astronomy}},
  \href{https://doi.org/10.3389/fspas.2018.00044}{\emph{Front. Astron. Space
  Sci.} {\bfseries 5} (2018) 44}
  [\href{https://arxiv.org/abs/1807.09241}{{\ttfamily 1807.09241}}].

\bibitem{Kase:2018aps}
R.~Kase and S.~Tsujikawa, \emph{{Dark energy in Horndeski theories after
  GW170817: A review}},  \href{https://arxiv.org/abs/1809.08735}{{\ttfamily
  1809.08735}}.

\bibitem{deRham:2018red}
C.~de~Rham and S.~Melville, \emph{{Gravitational Rainbows: LIGO and Dark Energy
  at its Cutoff}},
  \href{https://doi.org/10.1103/PhysRevLett.121.221101}{\emph{Phys. Rev. Lett.}
  {\bfseries 121} (2018) 221101}
  [\href{https://arxiv.org/abs/1806.09417}{{\ttfamily 1806.09417}}].

\bibitem{Horndeski:1974wa}
G.~W. Horndeski, \emph{{Second-order scalar-tensor field equations in a
  four-dimensional space}},
  \href{https://doi.org/10.1007/BF01807638}{\emph{Int. J. Theor. Phys.}
  {\bfseries 10} (1974) 363}.

\bibitem{Deffayet:2011gz}
C.~Deffayet, X.~Gao, D.~A. Steer and G.~Zahariade, \emph{{From k-essence to
  generalised Galileons}},
  \href{https://doi.org/10.1103/PhysRevD.84.064039}{\emph{Phys. Rev.}
  {\bfseries D84} (2011) 064039}
  [\href{https://arxiv.org/abs/1103.3260}{{\ttfamily 1103.3260}}].

\bibitem{Kobayashi:2011nu}
T.~Kobayashi, M.~Yamaguchi and J.~Yokoyama, \emph{{Generalized G-inflation:
  Inflation with the most general second-order field equations}},
  \href{https://doi.org/10.1143/PTP.126.511}{\emph{Prog. Theor. Phys.}
  {\bfseries 126} (2011) 511}
  [\href{https://arxiv.org/abs/1105.5723}{{\ttfamily 1105.5723}}].

\bibitem{Zumalacarregui:2013pma}
M.~Zumalac\'{a}rregui and J.~Garc\'{i}a-Bellido, \emph{{Transforming gravity:
  from derivative couplings to matter to second-order scalar-tensor theories
  beyond the Horndeski Lagrangian}},
  \href{https://doi.org/10.1103/PhysRevD.89.064046}{\emph{Phys. Rev.}
  {\bfseries D89} (2014) 064046}
  [\href{https://arxiv.org/abs/1308.4685}{{\ttfamily 1308.4685}}].

\bibitem{Gleyzes:2014dya}
J.~Gleyzes, D.~Langlois, F.~Piazza and F.~Vernizzi, \emph{{Healthy theories
  beyond Horndeski}},
  \href{https://doi.org/10.1103/PhysRevLett.114.211101}{\emph{Phys. Rev. Lett.}
  {\bfseries 114} (2015) 211101}
  [\href{https://arxiv.org/abs/1404.6495}{{\ttfamily 1404.6495}}].

\bibitem{Gleyzes:2014qga}
J.~Gleyzes, D.~Langlois, F.~Piazza and F.~Vernizzi, \emph{{Exploring
  gravitational theories beyond Horndeski}},
  \href{https://doi.org/10.1088/1475-7516/2015/02/018}{\emph{JCAP} {\bfseries
  1502} (2015) 018} [\href{https://arxiv.org/abs/1408.1952}{{\ttfamily
  1408.1952}}].

\bibitem{Langlois:2015cwa}
D.~Langlois and K.~Noui, \emph{{Degenerate higher derivative theories beyond
  Horndeski: evading the Ostrogradski instability}},
  \href{https://doi.org/10.1088/1475-7516/2016/02/034}{\emph{JCAP} {\bfseries
  1602} (2016) 034} [\href{https://arxiv.org/abs/1510.06930}{{\ttfamily
  1510.06930}}].

\bibitem{Langlois:2015skt}
D.~Langlois and K.~Noui, \emph{{Hamiltonian analysis of higher derivative
  scalar-tensor theories}},
  \href{https://doi.org/10.1088/1475-7516/2016/07/016}{\emph{JCAP} {\bfseries
  1607} (2016) 016} [\href{https://arxiv.org/abs/1512.06820}{{\ttfamily
  1512.06820}}].

\bibitem{Crisostomi:2016czh}
M.~Crisostomi, K.~Koyama and G.~Tasinato, \emph{{Extended Scalar-Tensor
  Theories of Gravity}},
  \href{https://doi.org/10.1088/1475-7516/2016/04/044}{\emph{JCAP} {\bfseries
  1604} (2016) 044} [\href{https://arxiv.org/abs/1602.03119}{{\ttfamily
  1602.03119}}].

\bibitem{Achour:2016rkg}
J.~Ben~Achour, D.~Langlois and K.~Noui, \emph{{Degenerate higher order
  scalar-tensor theories beyond Horndeski and disformal transformations}},
  \href{https://doi.org/10.1103/PhysRevD.93.124005}{\emph{Phys. Rev.}
  {\bfseries D93} (2016) 124005}
  [\href{https://arxiv.org/abs/1602.08398}{{\ttfamily 1602.08398}}].

\bibitem{BenAchour:2016fzp}
J.~Ben~Achour, M.~Crisostomi, K.~Koyama, D.~Langlois, K.~Noui and G.~Tasinato,
  \emph{{Degenerate higher order scalar-tensor theories beyond Horndeski up to
  cubic order}}, \href{https://doi.org/10.1007/JHEP12(2016)100}{\emph{JHEP}
  {\bfseries 12} (2016) 100}
  [\href{https://arxiv.org/abs/1608.08135}{{\ttfamily 1608.08135}}].

\bibitem{Langlois:2017mxy}
D.~Langlois, M.~Mancarella, K.~Noui and F.~Vernizzi, \emph{{Effective
  Description of Higher-Order Scalar-Tensor Theories}},
  \href{https://doi.org/10.1088/1475-7516/2017/05/033}{\emph{JCAP} {\bfseries
  1705} (2017) 033} [\href{https://arxiv.org/abs/1703.03797}{{\ttfamily
  1703.03797}}].

\bibitem{Langlois:2017mdk}
D.~Langlois, \emph{{Degenerate Higher-Order Scalar-Tensor (DHOST) theories}},
  in \emph{{Proceedings, 52nd Rencontres de Moriond on Gravitation (Moriond
  Gravitation 2017): La Thuile, Italy, March 25-April 1, 2017}}, pp.~221--228,
  2017, \href{https://arxiv.org/abs/1707.03625}{{\ttfamily 1707.03625}}.

\bibitem{Langlois:2018dxi}
D.~Langlois, \emph{{Dark Energy and Modified Gravity in Degenerate Higher-Order
  Scalar-Tensor (DHOST) theories: a review}},
  \href{https://arxiv.org/abs/1811.06271}{{\ttfamily 1811.06271}}.

\bibitem{Kobayashi:2019hrl}
T.~Kobayashi, \emph{{Horndeski theory and beyond: a review}},
  \href{https://arxiv.org/abs/1901.07183}{{\ttfamily 1901.07183}}.

\bibitem{Kobayashi:2014ida}
T.~Kobayashi, Y.~Watanabe and D.~Yamauchi, \emph{{Breaking of Vainshtein
  screening in scalar-tensor theories beyond Horndeski}},
  \href{https://doi.org/10.1103/PhysRevD.91.064013}{\emph{Phys. Rev.}
  {\bfseries D91} (2015) 064013}
  [\href{https://arxiv.org/abs/1411.4130}{{\ttfamily 1411.4130}}].

\bibitem{Crisostomi:2017lbg}
M.~Crisostomi and K.~Koyama, \emph{{Vainshtein mechanism after GW170817}},
  \href{https://doi.org/10.1103/PhysRevD.97.021301}{\emph{Phys. Rev.}
  {\bfseries D97} (2018) 021301}
  [\href{https://arxiv.org/abs/1711.06661}{{\ttfamily 1711.06661}}].

\bibitem{Langlois:2017dyl}
D.~Langlois, R.~Saito, D.~Yamauchi and K.~Noui, \emph{{Scalar-tensor theories
  and modified gravity in the wake of GW170817}},
  \href{https://doi.org/10.1103/PhysRevD.97.061501}{\emph{Phys. Rev.}
  {\bfseries D97} (2018) 061501}
  [\href{https://arxiv.org/abs/1711.07403}{{\ttfamily 1711.07403}}].

\bibitem{Dima:2017pwp}
A.~Dima and F.~Vernizzi, \emph{{Vainshtein Screening in Scalar-Tensor Theories
  before and after GW170817: Constraints on Theories beyond Horndeski}},
  \href{https://doi.org/10.1103/PhysRevD.97.101302}{\emph{Phys. Rev.}
  {\bfseries D97} (2018) 101302}
  [\href{https://arxiv.org/abs/1712.04731}{{\ttfamily 1712.04731}}].

\bibitem{Koyama:2015oma}
K.~Koyama and J.~Sakstein, \emph{{Astrophysical Probes of the Vainshtein
  Mechanism: Stars and Galaxies}},
  \href{https://doi.org/10.1103/PhysRevD.91.124066}{\emph{Phys. Rev.}
  {\bfseries D91} (2015) 124066}
  [\href{https://arxiv.org/abs/1502.06872}{{\ttfamily 1502.06872}}].

\bibitem{Saito:2015fza}
R.~Saito, D.~Yamauchi, S.~Mizuno, J.~Gleyzes and D.~Langlois, \emph{{Modified
  gravity inside astrophysical bodies}},
  \href{https://doi.org/10.1088/1475-7516/2015/06/008}{\emph{JCAP} {\bfseries
  1506} (2015) 008} [\href{https://arxiv.org/abs/1503.01448}{{\ttfamily
  1503.01448}}].

\bibitem{Sakstein:2015zoa}
J.~Sakstein, \emph{{Hydrogen Burning in Low Mass Stars Constrains Scalar-Tensor
  Theories of Gravity}},
  \href{https://doi.org/10.1103/PhysRevLett.115.201101}{\emph{Phys. Rev. Lett.}
  {\bfseries 115} (2015) 201101}
  [\href{https://arxiv.org/abs/1510.05964}{{\ttfamily 1510.05964}}].

\bibitem{Sakstein:2015aac}
J.~Sakstein, \emph{{Testing Gravity Using Dwarf Stars}},
  \href{https://doi.org/10.1103/PhysRevD.92.124045}{\emph{Phys. Rev.}
  {\bfseries D92} (2015) 124045}
  [\href{https://arxiv.org/abs/1511.01685}{{\ttfamily 1511.01685}}].

\bibitem{Jain:2015edg}
R.~K. Jain, C.~Kouvaris and N.~G. Nielsen, \emph{{White Dwarf Critical Tests
  for Modified Gravity}},
  \href{https://doi.org/10.1103/PhysRevLett.116.151103}{\emph{Phys. Rev. Lett.}
  {\bfseries 116} (2016) 151103}
  [\href{https://arxiv.org/abs/1512.05946}{{\ttfamily 1512.05946}}].

\bibitem{Sakstein:2016ggl}
J.~Sakstein, H.~Wilcox, D.~Bacon, K.~Koyama and R.~C. Nichol, \emph{{Testing
  Gravity Using Galaxy Clusters: New Constraints on Beyond Horndeski
  Theories}}, \href{https://doi.org/10.1088/1475-7516/2016/07/019}{\emph{JCAP}
  {\bfseries 1607} (2016) 019}
  [\href{https://arxiv.org/abs/1603.06368}{{\ttfamily 1603.06368}}].

\bibitem{Sakstein:2016lyj}
J.~Sakstein, M.~Kenna-Allison and K.~Koyama, \emph{{Stellar Pulsations in
  Beyond Horndeski Gravity Theories}},
  \href{https://doi.org/10.1088/1475-7516/2017/03/007}{\emph{JCAP} {\bfseries
  1703} (2017) 007} [\href{https://arxiv.org/abs/1611.01062}{{\ttfamily
  1611.01062}}].

\bibitem{Salzano:2017qac}
V.~Salzano, D.~F. Mota, S.~Capozziello and M.~Donahue, \emph{{Breaking the
  Vainshtein screening in clusters of galaxies}},
  \href{https://doi.org/10.1103/PhysRevD.95.044038}{\emph{Phys. Rev.}
  {\bfseries D95} (2017) 044038}
  [\href{https://arxiv.org/abs/1701.03517}{{\ttfamily 1701.03517}}].

\bibitem{Saltas:2018mxc}
I.~D. Saltas, I.~Sawicki and I.~Lopes, \emph{{White dwarfs and revelations}},
  \href{https://doi.org/10.1088/1475-7516/2018/05/028}{\emph{JCAP} {\bfseries
  1805} (2018) 028} [\href{https://arxiv.org/abs/1803.00541}{{\ttfamily
  1803.00541}}].

\bibitem{Babichev:2016jom}
E.~Babichev, K.~Koyama, D.~Langlois, R.~Saito and J.~Sakstein,
  \emph{{Relativistic Stars in Beyond Horndeski Theories}},
  \href{https://doi.org/10.1088/0264-9381/33/23/235014}{\emph{Class. Quant.
  Grav.} {\bfseries 33} (2016) 235014}
  [\href{https://arxiv.org/abs/1606.06627}{{\ttfamily 1606.06627}}].

\bibitem{Sakstein:2016oel}
J.~Sakstein, E.~Babichev, K.~Koyama, D.~Langlois and R.~Saito, \emph{{Towards
  Strong Field Tests of Beyond Horndeski Gravity Theories}},
  \href{https://doi.org/10.1103/PhysRevD.95.064013}{\emph{Phys. Rev.}
  {\bfseries D95} (2017) 064013}
  [\href{https://arxiv.org/abs/1612.04263}{{\ttfamily 1612.04263}}].

\bibitem{Chagoya:2018lmv}
J.~Chagoya and G.~Tasinato, \emph{{Compact objects in scalar-tensor theories
  after GW170817}},
  \href{https://doi.org/10.1088/1475-7516/2018/08/006}{\emph{JCAP} {\bfseries
  1808} (2018) 006} [\href{https://arxiv.org/abs/1803.07476}{{\ttfamily
  1803.07476}}].

\bibitem{Kobayashi:2018xvr}
T.~Kobayashi and T.~Hiramatsu, \emph{{Relativistic stars in degenerate
  higher-order scalar-tensor theories after GW170817}},
  \href{https://doi.org/10.1103/PhysRevD.97.104012}{\emph{Phys. Rev.}
  {\bfseries D97} (2018) 104012}
  [\href{https://arxiv.org/abs/1803.10510}{{\ttfamily 1803.10510}}].

\bibitem{Babichev:2018rfj}
E.~Babichev and A.~Leh\'{e}bel, \emph{{The sound of DHOST}},
  \href{https://doi.org/10.1088/1475-7516/2018/12/027}{\emph{JCAP} {\bfseries
  1812} (2018) 027} [\href{https://arxiv.org/abs/1810.09997}{{\ttfamily
  1810.09997}}].

\bibitem{Creminelli:2018xsv}
P.~Creminelli, M.~Lewandowski, G.~Tambalo and F.~Vernizzi, \emph{{Gravitational
  Wave Decay into Dark Energy}},
  \href{https://doi.org/10.1088/1475-7516/2018/12/025}{\emph{JCAP} {\bfseries
  1812} (2018) 025} [\href{https://arxiv.org/abs/1809.03484}{{\ttfamily
  1809.03484}}].

\bibitem{Frusciante:2018tvu}
N.~Frusciante, R.~Kase, K.~Koyama, S.~Tsujikawa and D.~Vernieri, \emph{{Tracker
  and scaling solutions in DHOST theories}},
  \href{https://doi.org/10.1016/j.physletb.2019.01.009}{\emph{Phys. Lett.}
  {\bfseries B790} (2019) 167}
  [\href{https://arxiv.org/abs/1812.05204}{{\ttfamily 1812.05204}}].

\bibitem{Koyama:2013paa}
K.~Koyama, G.~Niz and G.~Tasinato, \emph{{Effective theory for the Vainshtein
  mechanism from the Horndeski action}},
  \href{https://doi.org/10.1103/PhysRevD.88.021502}{\emph{Phys. Rev.}
  {\bfseries D88} (2013) 021502}
  [\href{https://arxiv.org/abs/1305.0279}{{\ttfamily 1305.0279}}].

\bibitem{Kimura:2011dc}
R.~Kimura, T.~Kobayashi and K.~Yamamoto, \emph{{Vainshtein screening in a
  cosmological background in the most general second-order scalar-tensor
  theory}}, \href{https://doi.org/10.1103/PhysRevD.85.024023}{\emph{Phys. Rev.}
  {\bfseries D85} (2012) 024023}
  [\href{https://arxiv.org/abs/1111.6749}{{\ttfamily 1111.6749}}].

\bibitem{Narikawa:2013pjr}
T.~Narikawa, T.~Kobayashi, D.~Yamauchi and R.~Saito, \emph{{Testing general
  scalar-tensor gravity and massive gravity with cluster lensing}},
  \href{https://doi.org/10.1103/PhysRevD.87.124006}{\emph{Phys. Rev.}
  {\bfseries D87} (2013) 124006}
  [\href{https://arxiv.org/abs/1302.2311}{{\ttfamily 1302.2311}}].

\bibitem{Terukina:2013eqa}
A.~Terukina, L.~Lombriser, K.~Yamamoto, D.~Bacon, K.~Koyama and R.~C. Nichol,
  \emph{{Testing chameleon gravity with the Coma cluster}},
  \href{https://doi.org/10.1088/1475-7516/2014/04/013}{\emph{JCAP} {\bfseries
  1404} (2014) 013} [\href{https://arxiv.org/abs/1312.5083}{{\ttfamily
  1312.5083}}].

\bibitem{Wilcox:2015kna}
H.~Wilcox et~al., \emph{{The XMM Cluster Survey: Testing chameleon gravity
  using the profiles of clusters}},
  \href{https://doi.org/10.1093/mnras/stv1366}{\emph{Mon. Not. Roy. Astron.
  Soc.} {\bfseries 452} (2015) 1171}
  [\href{https://arxiv.org/abs/1504.03937}{{\ttfamily 1504.03937}}].

\bibitem{Lopes:2003aa}
I.~P. Lopes and J.~Silk, \emph{{The sensitivity of the seismic solar model to
  Newton's constant}},
  \href{https://doi.org/10.1046/j.1365-8711.2003.06098.x}{\emph{Mon. Not. Roy.
  Astron. Soc.} {\bfseries 341} (2003) 721}.

\bibitem{Jimenez:2015bwa}
J.~Beltran~Jimenez, F.~Piazza and H.~Velten, \emph{{Evading the Vainshtein
  Mechanism with Anomalous Gravitational Wave Speed: Constraints on Modified
  Gravity from Binary Pulsars}},
  \href{https://doi.org/10.1103/PhysRevLett.116.061101}{\emph{Phys. Rev. Lett.}
  {\bfseries 116} (2016) 061101}
  [\href{https://arxiv.org/abs/1507.05047}{{\ttfamily 1507.05047}}].

\bibitem{deRham:2016wji}
C.~de~Rham and A.~Matas, \emph{{Ostrogradsky in Theories with Multiple
  Fields}}, \href{https://doi.org/10.1088/1475-7516/2016/06/041}{\emph{JCAP}
  {\bfseries 1606} (2016) 041}
  [\href{https://arxiv.org/abs/1604.08638}{{\ttfamily 1604.08638}}].

\end{thebibliography}\endgroup
\bibliographystyle{JHEP.bst}
\end{document}